\documentclass[manuscript]{acmart}

\usepackage{multirow}
\usepackage{siunitx}
\usepackage{subcaption}
\usepackage{braket}

\usepackage[ruled,vlined]{algorithm2e}

\AtBeginDocument{%
  }

\acmJournal{TQC}
\begin{document}

\title{\textsc{FlowQ-Net}: A Generative Framework for Automated Quantum Circuit Design}

\author{Jun Dai}
\email{jun.dai@mila.quebec}
\orcid{0000-0002-2732-7316}
\author{Michael Rizvi-Martel}
\author{Guillaume Rabusseau}
\affiliation{
  \institution{Mila - Quebec AI Institute, Université de Montréal}
  \city{Montréal}
  \state{QC}
  \country{Canada}
}
\affiliation{
  \institution{Département d’informatique et de recherche opérationnelle, Université de Montréal}
  \city{Montréal}
  \state{QC}
  \country{Canada}
}

\renewcommand{\shortauthors}{Dai et al.}

\begin{abstract}
Designing efficient quantum circuits is a central bottleneck to exploring the potential of quantum computing, particularly for noisy intermediate-scale quantum (NISQ) devices, where circuit efficiency and resilience to errors are paramount. 
The search space of gate sequences grows combinatorially, and handcrafted templates often waste scarce qubit and depth budgets. 
We introduce \textsc{FlowQ-Net} (Flow-based Quantum design Network), a generative framework for automated quantum circuit synthesis based on Generative Flow Networks (GFlowNets). 
This framework learns a stochastic policy to construct circuits sequentially, sampling them in proportion to a flexible, user-defined reward function that can encode multiple design objectives such as performance, depth, and gate count. This approach uniquely enables the generation of a diverse ensemble of high-quality circuits, moving beyond single-solution optimization. We demonstrate the efficacy of \textsc{FlowQ-Net} through an extensive set of simulations.
We apply our method to Variational Quantum Algorithm (VQA) ansatz design for molecular ground state estimation, Max-Cut, and image classification, key challenges in near-term quantum computing.
Circuits designed by \textsc{FlowQ-Net} achieve significant improvements, yielding circuits that are 10$\times$-30$\times$ more compact in terms of parameters, gates, and depth compared to commonly used unitary baselines, without compromising accuracy.
This trend holds even when subjected to error profiles from real-world quantum devices.
Our results underline the potential of generative models as a general-purpose methodology for automated quantum circuit design, offering a promising path towards more efficient quantum algorithms and accelerating scientific discovery in the quantum domain.
\end{abstract}




\maketitle

\section{Introduction}

Quantum circuits are the fundamental primitives of quantum computation, yet designing quantum circuits represents a significant challenge. 
In the current Noisy Intermediate-Scale Quantum (NISQ) era \cite{preskill2018quantum,lau2022nisq}, devices suffer from limitations, including modest qubit numbers, restricted connectivity, finite coherence times, and noise. 
This makes circuit depth and gate count critical factors for determining success \cite{chow2021ibm}.
While manual design can handle small systems, human intuition fails to navigate the exponentially growing optimization landscape of larger architectures, often yielding unnecessarily deep or over-parameterized circuits that exacerbate error accumulation. These challenges motivate the development of automated circuit-design methods applicable to diverse goals such as algorithm compilation, quantum-state preparation, and parameterized ansatz construction for Variational Quantum Algorithms (VQAs)~\cite{farhi2014quantum,peruzzo2014variational}.

The challenge stems from the exponential growth of the search space for optimal gate sequences, which makes identifying compact yet expressive circuits computationally intractable. NISQ hardware further constrains feasible designs, demanding circuits that balance expressivity and resource efficiency. In quantum chemistry, for instance, the widely used unitary coupled-cluster ansatz with singles and doubles (UCCSD)~\cite{romero2018strategies,lee2018generalized} involves numerous excitation operators, producing circuits too deep for near-term devices. Conversely, hardware-efficient ansätze~\cite{kandala2017hardware} yield shallower structures but often lack sufficient expressivity to achieve accurate results.

To overcome these challenges, a variety of automated approaches have been developed to construct efficient variational quantum circuits. Adaptive VQE methods~\cite{grimsley2019adaptive,sambasivam2025tepid} iteratively build an ansatz by adding gate operators that most improve the target energy, thereby tailoring circuit structure to the underlying problem. Reinforcement learning (RL) approaches~\cite{haarnoja2017reinforcement,ostaszewski2021reinforcement,patel2024curriculum,ye2021quantum,kuo2021quantum} have also been used to sequentially grow quantum circuits; for instance, Patel \textit{et al.}~\cite{patel2024curriculum} trained an RL agent to generate shallow, high-performance circuits for LiH. Beside RL, evolutionary and differentiable methods~\cite{massey2004evolving,zhang2022differentiable,wu2023quantumdarts} search for circuit architectures through either population-based heuristics or gradient-driven relaxations of discrete gate choices. The Generative Quantum Eigensolver (GQE)~\cite{nakaji2024generative} and its transformer-based variant GPT-QE use pre-trained generative models to synthesize circuits for molecular Hamiltonians, while recent diffusion-based approaches~\cite{furrutter2024quantum} extend this paradigm to unitary compilation and entanglement generation. Other frameworks, such as Bayesian optimization~\cite{duong2022quantum} and Monte Carlo Tree Search (MCTS)~\cite{lipardi2025quantum}, have also been proposed to improve circuit search efficiency.

Despite these advances, existing Quantum Architecture Search strategies share fundamental limitations. RL and MCTS methods often suffer from sample inefficiency, instability, or convergence to a single high-reward solution~\cite{li2020quantum,ostaszewski2021reinforcement,patel2024curriculum}, while evolutionary algorithms lack formal guarantees and may prematurely stagnate~\cite{massey2004evolving,peruzzo2014variational}. Differentiable approaches such as DQAS~\cite{zhang2022differentiable,wu2023quantumdarts} offer gradient-based optimization but incur high computational costs and rely on continuous relaxations that obscure discrete circuit structures. Consequently, systematically exploring the exponentially large and structured circuit space remains an open challenge. Most current methods aim to identify a single optimized circuit rather than to learn a generative process capable of producing a diverse portfolio of compact, high-quality designs that balance expressivity, resource efficiency, and robustness under realistic noise constraints.

As an alternative to conventional search strategies, Generative Flow Networks (GFlowNets) \cite{bengio2021flow,zahavy2021reward,zhou2023phylogfn,nishikawa2022bayesian,hu2023amortizing,grathwohl2021oops,dai2020learning,buesing2020approximate,tiapkin2024generative,malkin2022gflownets} offer a promising alternative for circuit design.
GFlowNets learn a stochastic policy through a flow-matching objective that enforces consistency between forward flows, propagating probability mass from initial to terminal states, and backward flows in the reverse direction.
This training principle ensures that complete objects are sampled in proportion to their reward, yielding a normalized reward distribution~\cite{bengio2023gflownet} that naturally encourages both diversity and exploration.
These properties make GFlowNets particularly suitable for the combinatorial nature of quantum circuit design, where a vast space of gate sequences must be explored efficiently.

In this work, we present \textsc{FlowQ-Net} (Flow-based Quantum design Network), a generative framework for automated quantum circuit synthesis built upon the GFlowNet paradigm~\cite{bengio2021flow}. \textsc{FlowQ-Net} formulates the circuit search as a sequential decision process in which a GFlowNet learns to sample candidate architectures in proportion to their performance. This formulation naturally extends to quantum circuits, which are structured, compositional objects composed of sequential gate operations. By learning the entire distribution of high-quality circuits rather than converging on a single optimal design, \textsc{FlowQ-Net} explicitly promotes diversity in the generated architectures. This diversity allows it to capture multiple modes in complex reward landscapes, yielding a richer characterization of the design space and a broader repertoire of competitive solutions. Empirical results demonstrate that \textsc{FlowQ-Net} achieves substantial performance gains under NISQ constraints, producing compact and accurate circuits while maintaining scalability. Furthermore, its inherently parallel generative process enables efficient exploration of large design spaces, where heuristic pruning can be employed to further accelerate discovery.

\section{Methods}
\subsection{Variational Quantum Algorithm (VQA)}
In VQA, a quantum circuit
$U(\boldsymbol{\theta};\mathcal{G})$,often called an \emph{ansatz},is parameterized by a
set of real parameters $\boldsymbol{\theta}$ and a discrete architecture
$\mathcal{G}$ describing its gate sequence and connectivity. The circuit prepares a
parameterized quantum state
\[
|\psi(\boldsymbol{\theta};\mathcal{G})\rangle = U(\boldsymbol{\theta};\mathcal{G})|\psi_0\rangle,
\]
where $|\psi_0\rangle$ is an easily preparable initial state, typically the all‐zero
state.

A classical optimizer iteratively updates $\boldsymbol{\theta}$ to minimize a task‐specific
loss function $\mathcal{L}(\boldsymbol{\theta};\mathcal{G})$, which depends on measurements
obtained from the quantum device. For example, in the VQE, the objective is to minimize the expectation value of a molecular Hamiltonian
$H$,
\[
\mathcal{L}(\boldsymbol{\theta};\mathcal{G}) =
\langle\psi(\boldsymbol{\theta};\mathcal{G})|\,H\,|\psi(\boldsymbol{\theta};\mathcal{G})\rangle,
\]
whose minimum approximates the ground‐state energy.
Other instances include the optimization problems,
where $\mathcal{L}$ encodes a cost Hamiltonian for combinatorial optimization,
and Quantum Neural Networks (QNNs), where $\mathcal{L}$ is a supervised
classification loss such as cross‐entropy.

The quality of a VQA solution depends critically on the expressive power and trainability
of the chosen ansatz $\mathcal{G}$. Conventional designs,such as the hardware‐efficient
ansatz or chemically motivated constructions like UCCSD, offer fixed circuit structures
that can be suboptimal for a given problem instance. Designing efficient and
problem‐adapted ansätze therefore becomes an architectural search problem, one that
rapidly grows combinatorially with system size and hardware constraints.

In the FlowQ‐Net framework introduced here, the continuous optimization over
$\boldsymbol{\theta}$ remains a classical inner loop, while the discrete structure
$\mathcal{G}$ is treated as the object of a generative search handled by the outer
GFlowNet. This separation yields a bi‐level optimization scheme: the inner variational
loop evaluates $\mathcal{L}(\boldsymbol{\theta};\mathcal{G})$ for candidate circuits,
and the outer loop learns to sample promising architectures in proportion to their
resulting reward $R(\mathcal{G})$. Such integration enables FlowQ‐Net to inherit
the interpretability and robustness of VQAs while automating their structural design.

\subsection{GFlowNets}
\label{sec:gflownet}

GFlowNets ~\cite{bengio2021flow,bengio2023gflownet}
are a class of probabilistic generative models designed to learn 
\emph{stochastic policies} that sample compositional objects $x$ (e.g., 
molecular graphs, quantum circuits, or symbolic expressions) 
in proportion to a user-defined reward function $R(x) > 0$.
Unlike standard reinforcement learning (RL) approaches, which optimize
for a single high-reward trajectory, GFlowNets aim to learn a distribution
over all possible high-reward objects, balancing exploration and diversity.

A GFlowNet defines a directed acyclic graph $G =(\mathcal{S}, \mathcal{A})$
where nodes $s \in \mathcal{S}$ represent \emph{states}
and edges $(s \!\to\! s') \in \mathcal{A}$ correspond to \emph{actions} that extend a state.
The generation process starts from an empty initial state $s_0$ and proceeds
through a sequence of actions $(s_0 \!\to\! s_1 \!\to\! \cdots \!\to\! s_T)$ until a terminal state
$s_T$ is reached, representing a complete object $x = s_T$.

The model maintains two stochastic policies:
the \emph{forward policy} $P_F(s'|s)$, which samples the next state during construction,
and the \emph{backward policy} $P_B(s|s')$, which allows learning via flow consistency.
The goal of training is to make the expected \emph{flow of probability mass}
through each state consistent between the forward and backward directions:
\begin{equation}
    F(s) = \!\!\sum_{s' \in \text{parent}(s)} \! P_F(s|s')F(s')
          = \!\!\sum_{s' \in \text{child}(s)} \! P_B(s'|s)F(s'),
\end{equation}
where $F(s)$ denotes the total incoming flow at state $s$.
At terminal states $s_T$, this flow is constrained by
\begin{equation}
    F(s_T) = R(s_T),
\end{equation}
which ensures that the probability of sampling a complete object $x$
is proportional to its reward:
\[
P(x) \propto R(x).
\]

Several equivalent loss formulations exist for training GFlowNets.
A common objective is the \emph{Trajectory Balance} (TB) loss~\cite{malkin2022gflownets},
which directly enforces global flow consistency across entire trajectories,
\begin{equation}
    \mathcal{L}_{\text{TB}}(\tau) =
    \Bigl(
        \log Z_\phi
        + \sum_{(s \to s') \in \tau}\!\log P_F(s'|s)
        - \log R(x)
        - \!\!\sum_{(s' \to s) \in \tau}\!\log P_B(s|s')
    \Bigr)^2.
    \label{eq:tb}
\end{equation}
Here, $\tau$ is a sampled trajectory ending in terminal state $x=s_T$,
$Z_\phi$ is a learned normalizing constant (analogous to a partition function),
and $(P_F, P_B)$ are parameterized by neural networks.
Minimizing Eq.~\eqref{eq:tb} ensures that the forward and backward flows 
match up to a constant scaling factor $Z_\phi$, guaranteeing a valid normalized distribution.

In the context of quantum computing, the space of valid circuits is 
combinatorial and highly structured. Each circuit can be seen as a trajectory
of discrete design decisions (gate placement, termination).
The flow-matching principle allows GFlowNets to explore this structured space 
efficiently while maintaining diversity among generated architectures.
When integrated into the bi-level optimization of a VQA,
this yields the central mechanism behind our \textsc{FlowQ-Net} framework
(Section~\ref{sec:method}), where the GFlowNet learns to sample 
ansatz structures proportional to their performance-derived rewards.

\subsection{\textsc{FlowQ-Net} for Quantum Circuit Design}
\label{sec:method}

\subsubsection{Problem formulation}

\begin{figure}[t]
  \centering
  \includegraphics[width=0.8\linewidth]{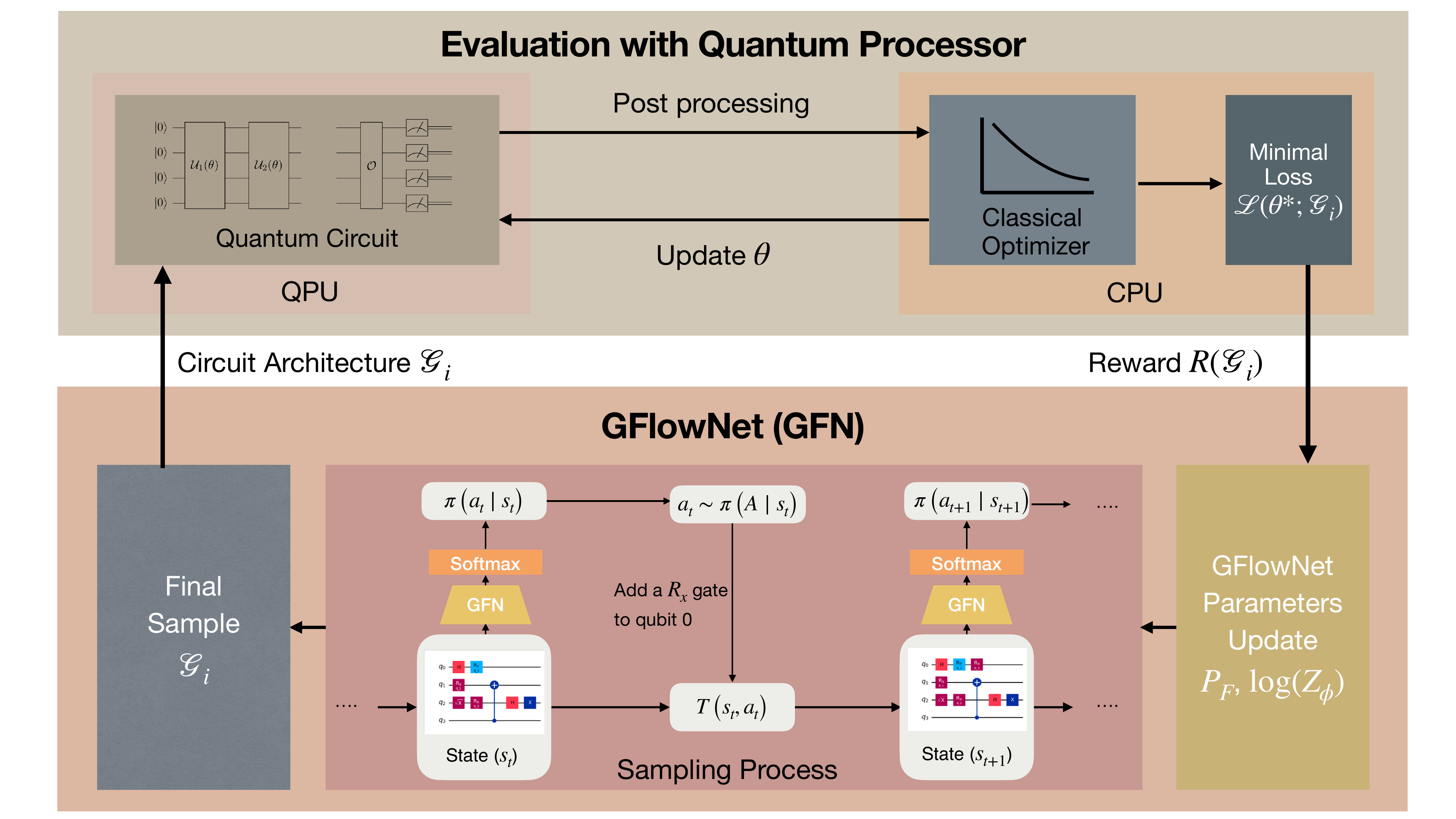}
  \caption{\textbf{GFlowNet-guided VQA ansatz design.} 
  Top band: For each candidate architecture $\mathcal{G}_i$ generated by \textsc{FlowQ-Net}, the quantum‐processing unit (QPU) executes the circuits to obtain measurement outcomes, which are post‐processed to obtain a loss. If parameters are present, a classical optimizer optimizes continuous parameters $\boldsymbol{\theta}$ to minimize the loss $\mathcal{L}(\boldsymbol{\theta};\mathcal{G}_i)$, while parameter-free circuits are evaluated directly.  
  The process returns the minimal loss $\mathcal{L}(\boldsymbol{\theta}^*;\mathcal{G}_i)$. Bottom band: The resulting loss is mapped to a reward $R(\mathcal{G}_i)$, which updates the GFlowNet component of \textsc{FlowQ-Net}.
  then a new circuit architecture  $\mathcal{G}_i$ is constructed step by step by sampling actions $a_t\!\sim\!\pi(a_t\mid s_t)$ from the GFlowNet.  The terminal sample is finally passed to the QPU.
  %, and the reward $R(\mathcal{G}_i)$ that drives a update of the Flow model parameters $P_F$ and the log-partition scale $\log Z_{\phi}$. 
  The outer GFlowNet learning loop and the inner (optional) continuous parameter optimization loop iterate until high-reward (low-loss) circuit architectures are sampled with high probability.}
  \label{fig:workflow}
\end{figure}

\textsc{FlowQ-Net} formulates \emph{quantum-circuit synthesis} as a sequential generative process. 
At the core of \textsc{FlowQ-Net}, a GFlowNet learns to sample in proportion to a user-defined \emph{reward} $R(x)$.
Fig \ref{fig:workflow} summarizes our \textsc{FlowQ-Net} approach. 
In the following, we give a parameterization of quantum-circuit synthesis as a RL decision problem.  
The formulation is intentionally task–agnostic so that, by simply re-defining $R(x)$, the \emph{same} learning algorithm can target ansatz discovery, compilation of a target unitary and state preparations in a unified way and can easily be extended to other tasks.

\paragraph{State space.}
A state $s_t$ is a \emph{partial circuit}, i.e.\ an ordered list of quantum
gates that have been placed so far; the empty list $s_0=\varnothing$ is the initial state. 

\paragraph{Action.}
At every step the agent selects an action $a_t \in \mathcal{A}(s_t)$ that either
 appends a gate from a gate set $\mathbf{G}$, e.g. $\{R_x,R_y,R_z,\textsc{CNOT}\}$ on its target qubits, or outputs \textsc{stop} to terminate the episode. The gate set $\mathbf{G}$ is universal  \cite{nielsen2001quantum}. The action mask enforces simple syntactic validity.

\paragraph{Reward.} On termination, the completed circuit $x=s_T$ is evaluated to obtain a
scalar reward $R(x)$ defined by
\begin{equation}
    \label{eq:reward}
     \log R(x)= -\beta\;\bigl[\min_{\boldsymbol{\theta}}\mathcal{L}(x,\boldsymbol{\theta}) - b\bigr],
\end{equation}
 
where $\beta$ controls exploration and $b$ is an offset. The loss $\mathcal{L}$ is task-dependent, e.g., variational energy for ground-state estimation %the negative cut value for Max-Cut, 
and cross-entropy for QNN classification. This \emph{task-aware} shaping steepens the landscape around chemically accurate or combinatorially optimal circuits, guiding the policy towards compact, high-quality ansatz. The loss value is computed by fitting circuit parameters and computing the associated loss; this step is done on the quantum device directly. 

\paragraph{Parameterized gates.} Within \textsc{FlowQ-Net}, the GFlowNet decides \emph{only} the discrete structure; each candidate circuit’s continuous parameters $\boldsymbol{\theta}$ are optimized classically with gradient based optimizer to obtain $\mathcal{L}(\boldsymbol{\theta}^*;\mathcal{G})$. Although computationally intensive, this bi-level design (structure search outer loop, parameter optimization inner loop) yields a faithful reward signal and is crucial for objectives such as VQA.

\subsubsection{GFlowNet training}
To train a GFlowNet, the first step is to define a \textit{flow function}. Though many such functions exist, we opt for the trajectory balance loss (Eq. \ref{eq:tb}), as it is known to lead to better credit assignment~\cite{malkin2022trajectory}. Here, $P_F$ represents the forward policy, $P_B$ the backwards policy and $Z_\phi$ is the normalizing constant for the reward distribution.
Typically, $P_F(a\mid s;\phi)$ is implemented using a neural network and $Z_\phi$ is set to be a learnable scalar parameter. We set $P_B$ to be a uniform distribution over previous states as this reduces the number of estimators to learn and empirically helps with convergence.

\paragraph{Optimization loop.}
The resulting circuits are evaluated to compute their rewards and trajectory-balance losses, after which the Transformer parameters and the partition scalar $\log Z_\phi$ are updated via Adam.  
A high inverse temperature coupled with an adaptive baseline concentrates probability mass on circuits already achieving chemical accuracy or optimal cuts while preserving diversity across the search space (see Appendix~\ref{app:imple} for full algorithmic details). 

\paragraph{Computational Complexity Compared to RL and Differentiable QAS}
A key advantage of the proposed \textsc{FlowQ-Net} framework lies in its computational efficiency.  
Because training relies on terminal rewards, the cost of one optimization round scales linearly with the number of complete circuit evaluations, $\mathcal{O}(N)$, where $N$ is the number of trajectories (circuits) sampled per update.  
In contrast, reinforcement learning based quantum architecture search requires per-step reward evaluations for each gate addition, resulting in a step-wise cost of $\mathcal{O}(N \times \text{depth})$.  
Differentiable quantum architecture search methods such as QuantumDARTS~\cite{wu2023quantumdarts} incur a higher complexity due to continuous relaxation and gradient computations through parameterized unitaries, with a reported cost of $\mathcal{O}(4^n N d)$ for circuits of depth $d$ and number of qubit $n$.  

Once trained, \textsc{FlowQ-Net} performs inference purely through classical sampling from the learned forward policy $P_F(a|s)$ without any quantum evaluations or gradient updates.  
This makes architecture generation highly efficient and trivially parallelizable across CPUs or GPUs, providing a substantial runtime advantage for large-scale circuit synthesis tasks.

\section{Experiments}
\label{sec:exp}
Our objective is to determine whether the \textsc{FlowQ-Net} framework of
Sec.~\ref{sec:method} can discover compact, noise-robust ansatz that match or
surpass the accuracy of established baselines while using fewer quantum circuit resources. 
%Given the prohibitive environmental footprint~\cite{martin2022energy,cordier2025scaling} and restrained access of real quantum devices, 
All our experiments are performed via simulation using the PennyLane platform~\cite{bergholm2018pennylane}. In order to test the viability of our method in a NISQ setting, we conduct further experiments using noise profiles from real-word quantum computers. Note that the use of simulations is common in practice; many state of the art quantum algorithms validate their results through simulation~\cite{caro2022generalization,patel2024curriculum,wu2023quantumdarts,wan2022randomized,furrutter2024quantum}.

For the GFlowNet implementation, we use a transformer network with a hidden size of 2 , 048, 3 layers, an embedding dimension of 512, and 8 attention heads for the forward policy. The backward policy is modeled by a uniform distribution to quickly approach one of the optimal solutions as suggested in~\cite{malkin2022trajectory}. We implement all our models in PyTorch \cite{paszke2019pytorch}, and we adopt trajectory balance as the main objective function. The model is updated every 5 iterations, after computing the batch loss for 5 complete trajectories. We use the Adam optimizer with a learning rate of $1 \times 10^{-4}$ for the parameters in our models. The inverse temperature parameter $\beta$ used in the reward function, is set to 1. All quantum computations are performed using simulators provided by PennyLane, a comprehensive software framework for quantum computing \cite{bergholm2018pennylane}. We leverage PennyLane’s state vector simulator to execute the quantum circuits. The circuit parameters $\theta$ are optimized using PennyLane’s built-in Adam optimizer, with a learning rate of $1 \times 10^{-2}$, 5 random reinitializations, and a maximum of 500 optimization steps.

\subsection{Quantum chemistry benchmarks}
\label{sec:exp_chem}

One of the key applications of the VQA framework is the Variational Quantum Eigensolver (VQE) \cite{peruzzo2014variational}, which is designed to find the ground state energy, or lowest energy, $E_0$, of a given Hamiltonian $\mathcal{H}$. This task is at the heart of quantum chemistry, as knowing the ground state energy of a molecule enables accurate predictions of its stability, reactivity, and electronic properties.
VQE has its roots in the variational principle of quantum mechanics, which guarantees that the expectation value of the Hamiltonian $\mathcal{H}$, calculated with any trial quantum state $|\psi(\boldsymbol{\theta}) \rangle = U(\boldsymbol{\theta};\mathcal{G}) |\psi_0\rangle$, will always be lower bounded by the ground state energy $E_0$. 

\begin{table}[t]
\centering
\small
\setlength{\tabcolsep}{5pt}
\renewcommand{\arraystretch}{1.1}
\caption{Atomic coordinates of molecules used in the VQE experiments.
All distances are given in Ångströms (Å).}
\label{tab:geom}
\vspace{0.4em}

\begin{tabular}{llccc}
\toprule
\textbf{Molecule} & \textbf{Qubit encoding} & \textbf{Atom} & \textbf{Coordinates (Å)} & \textbf{Bond length (Å)} \\
\midrule
H$_2$ & 4q &
\begin{tabular}[c]{@{}l@{}}H$_1$\\H$_2$\end{tabular} &
\begin{tabular}[c]{@{}l@{}}(0.000, 0.000, 0.000)\\(0.000, 0.000, 0.7414)\end{tabular} & 0.7414 \\
\midrule
LiH & 4q &
\begin{tabular}[c]{@{}l@{}}Li\\H\end{tabular} &
\begin{tabular}[c]{@{}l@{}}(0.000, 0.000, 0.000)\\(0.000, 0.000, 3.400)\end{tabular} & 3.4 \\
LiH & 6q &
\begin{tabular}[c]{@{}l@{}}Li\\H\end{tabular} &
\begin{tabular}[c]{@{}l@{}}(0.000, 0.000, 0.000)\\(0.000, 0.000, 2.200)\end{tabular} & 2.2 \\
\midrule
H$_2$O & 8q &
\begin{tabular}[c]{@{}l@{}}O\\H$_1$\\H$_2$\end{tabular} &
\begin{tabular}[c]{@{}l@{}}(0.835, 0.452, 0.000)\\(-0.021, -0.002, 0.000)\\(1.477, -0.273, 0.000)\end{tabular} & -- \\
\bottomrule
\end{tabular}
\end{table}

We evaluate ground-state energy estimation for \mbox{H$_2$}, LiH, and \mbox{H$_2$O} in the STO-3G basis. 
Electronic Hamiltonians are mapped to qubits via the Jordan–Wigner transformation \citep{jordan1993paulische}, 
except for the LiH–4q instance, for which we employ the parity mapping.  Here, the suffix “mq” signifies that the corresponding molecular problem is encoded on $m$ qubits.
Our study therefore comprises four problem sizes: \mbox{H$_2$–4q}, LiH–4q, LiH–6q, and \mbox{H$_2$O–8q}  
(see Table \ref{tab:geom}). We compare \textsc{FlowQ-Net} against four established ansatz-construction methods,  
(i) UCCSD \cite{lee2018generalized},  
(ii) ADAPT-VQE \citep{grimsley2019adaptive},
(iii) QuantumDARTS\cite{wu2023quantumdarts}
and   (iv) CRLQAS\cite{patel2024curriculum}.  

To mimick the conditions of a real-world quantum device, we also test \textsc{FlowQ-Net} using calibrated error profiles. As these noisy simulations are computationally demanding, we only run these experiments on smaller molecules \mbox{H$_2$–2q}, \mbox{H$_2$–3q}, \mbox{H$_2$–4q}, and \mbox{LiH–4q}.

\subsubsection{Noiseless simulation results}
\label{subsec:noiseless_results}

\begin{table}[t]
\centering
\small
\setlength{\tabcolsep}{4pt}
\renewcommand{\arraystretch}{1.05}
\caption{Comparison of circuit quality (\textit{energy difference} $E$, \textit{number of parameters} $P$, \textit{circuit depth} $D$, total \textit{gate count} $G$, and \textit{CNOT count} $C$) on four molecular VQE benchmarks. Boldface indicates the best (lowest) value for each metric within a molecule. ``--'' means no improvement over Hartree–Fock energy.}
\label{tab:vqe_main}
\vspace{0.4em}

\resizebox{0.5\linewidth}{!}{%
\begin{tabular}{llccccc}
\toprule
\textbf{Molecule} & \textbf{Method} & $E$ & $P\!\downarrow$ & $D\!\downarrow$ & $G\!\downarrow$ & $C\!\downarrow$ \\
\midrule
\multirow{5}{*}{H$_2$-4q}
 & \textsc{FlowQ-Net} (Ours) & 1.6e$^{-15}$ & \textbf{3} & \textbf{10} & \textbf{16} & \textbf{13} \\
 & UCCSD & 5.5e$^{-11}$ & 53 & 74 & 100 & 47 \\
 & APAPT-VQE & 1.9e$^{-2}$ & 38 & 29 & 38 & 0 \\
 & QuantumDARTS & 4.3e$^{-6}$ & 26 & 18 & 34 & 8 \\
 & CRLQAS & 7.2e$^{-8}$ & 7 & 17 & 21 & 14 \\
\midrule
\multirow{5}{*}{LiH-4q}
 & \textsc{FlowQ-Net} (Ours) & 2.0e$^{-5}$ & \textbf{8} & 32 & 43 & \textbf{35} \\
 & UCCSD & 4.0e$^{-5}$ & 159 & 223 & 309 & 150 \\
 & APAPT-VQE & 4.6e$^{-6}$ & 47 & 38 & 47 & 0 \\
 & QuantumDARTS & 1.7e$^{-4}$ & 50 & 34 & 68 & 18 \\
 & CRLQAS & 2.6e$^{-6}$ & 29 & \textbf{22} & \textbf{40} & 11 \\
\midrule
\multirow{5}{*}{LiH-6q}
 & \textsc{FlowQ-Net} (Ours) & 7.2e$^{-4}$ & \textbf{20} & \textbf{30} & \textbf{41} & \textbf{21} \\
 & UCCSD & 4.0e$^{-5}$ & 224 & 347 & 464 & 240 \\
 & APAPT-VQE & -- & -- & -- & -- & -- \\
 & QuantumDARTS & 2.9e$^{-4}$ & 80 & 54 & 132 & 52 \\
 & CRLQAS & 6.7e$^{-4}$ & 29 & 40 & 67 & 38 \\
\midrule
\multirow{5}{*}{H$_2$O-8q}
 & \textsc{FlowQ-Net} (Ours) & 5.2e$^{-4}$ & 50 & \textbf{38} & \textbf{107} & \textbf{57} \\
 & UCCSD & 4.0e$^{-6}$ & 962 & 1705 & 2180 & 1218 \\
 & APAPT-VQE & 2.6e$^{-3}$ & -- & -- & -- & -- \\
 & QuantumDARTS & 3.1e$^{-4}$ & 151 & 64 & 219 & 68 \\
 & CRLQAS & 1.8e$^{-4}$ & \textbf{35} & 75 & 140 & 105 \\
\bottomrule
\end{tabular}}
\end{table}

Table~\ref{tab:vqe_main} summarizes the circuit quality obtained by our \textsc{FlowQ-Net} framework in noise–free simulators, compared against four strong baselines. For all considered tasks, our method recovers the ground state energy up to the commonly accepted chemical accuracy of $1.6{\times}10^{-3}$~Ha.
For the smallest system (H$_2$–4q) our method finds the ground–state energy at least four orders of magnitude less parameters than the considered benchmarks. The underlying structure and expressivity of the learned ansatz are further elucidated through a Dynamical Lie Algebra analysis (\cite{schirmer2002identification,diaz2023showcasing}, see Section \ref{sec:dla}), which quantifies the minimal algebra generated by the circuit and its correspondence to the target molecular subspace.
On LiH–4q the best absolute error is achieved by CRLQAS ($2.6{\times}10^{-6}$ Ha), yet \textsc{FlowQ-Net} is within a factor of~8 while using 3.6$\times$ fewer parameters.  
For the larger LiH–6q and H$_2$O–8q tasks our energies remain within $7.2{\times}10^{-4}$~Ha and $5.2{\times}10^{-4}$~Ha respectively, well inside the chemical accuracy, despite radically reduced circuit resources.

Across all benchmarks \textsc{FlowQ-Net} generates the most compact circuits. On H$_2$–4q only three variational parameters are required, $17\times$ fewer than QuantumDARTS and $18\times$ fewer than UCCSD, substantially reducing classical optimisation cost.
Even for the eight-qubit H$_2$O instance our 50-parameter circuit is an order of magnitude smaller than UCCSD (962 parameters) and comparable with CRLQAS results. 

Circuit depth~$D$ and total gate count~$G$ are the dominant cost drivers on NISQ hardware.  
\textsc{FlowQ-Net} achieves the shallowest depth on every molecule except LiH–4q, where CRLQAS edges ahead by ten layers at the cost of 3.6$\times$ more parameters. Notably, for H$_2$O–8q we compress UCCSD’s $1\,705$-layer circuit to just \textbf{38} layers and reduce the gate count from $2\,180$ to \textbf{107} (a $20\times$ reduction) without leaving the chemical-accuracy regime.

From four to eight qubits, baseline resource metrics grow super-linearly, UCCSD’s parameter count explodes from 53 to 962 ($18\times$).  In contrast, \textsc{FlowQ-Net} scales much more gently (3 $\rightarrow$ 50 parameters; depth 10 $\rightarrow$ 38), suggesting that the generative search discovers reusable structural motifs that generalize with system size.  
This amortized advantage becomes increasingly pronounced as problem dimensionality grows.

\begin{figure}[h]
  \centering

    \includegraphics[width=\linewidth]{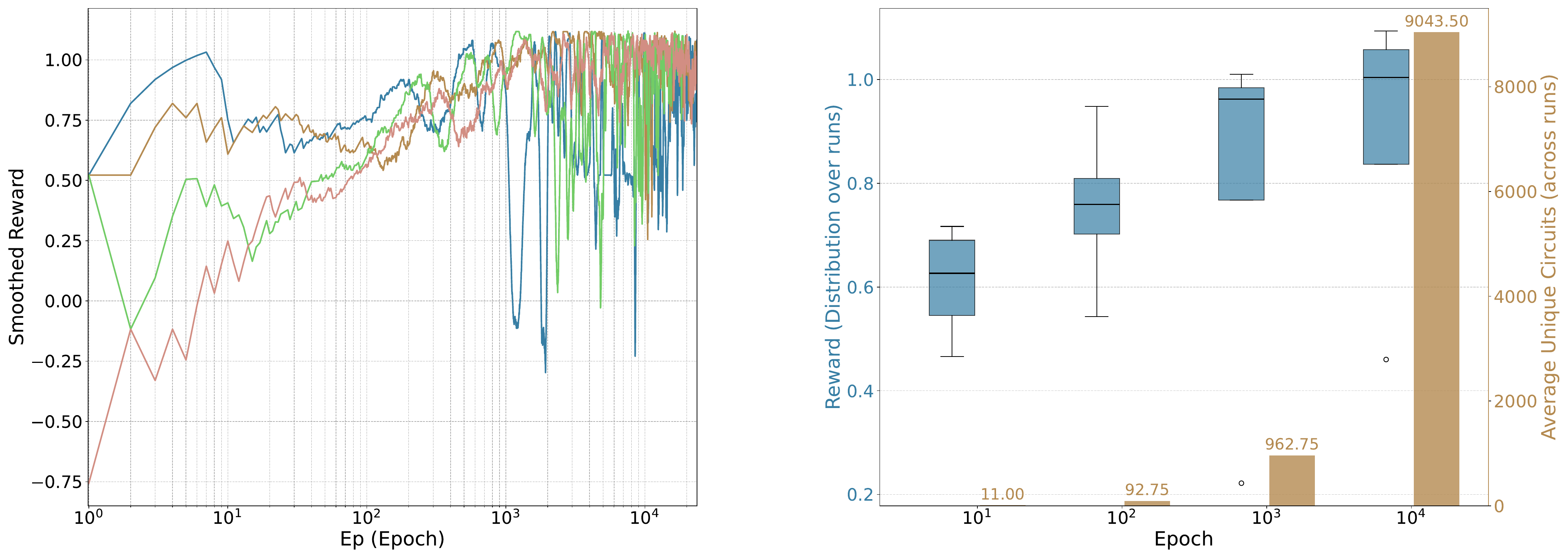}

  \caption{\textbf{Learning dynamics of the \textsc{FlowQ-Net} on four independent \(\mathrm{H}_2\) (4-qubit) runs.} Left panel shows the reward approaching its theoretical optimum after roughly \(10^{3}\) epochs. A running mean over 50 epochs smooths the raw reward. Right panel shows a steady expansion in circuit diversity.  The two plots indicate that exploration and exploitation progress yielding high-quality and diverse circuit families.}
  \label{fig:gfn_learning_dynamics}
\end{figure}

Figure \ref{fig:gfn_learning_dynamics} explains the optimization behavior and supports the results in Table \ref{tab:vqe_main}. In the right panel, the mean reward rises sharply during the first $10^3$ epochs and the model keep sample high reward (low loss) candidate for the H$_2$-4q task. The left panel shows that the \textsc{FlowQ-Net} continues to discover new circuit architectures when the mean rewards reach the highest value. This explain why our parameter count and depth scale sub-linearly from four to eight qubits.

The noiseless simulations demonstrate that a generative, reward-proportional exploration strategy can simultaneously \emph{minimize quantum resources} and \emph{preserve chemical accuracy}.  
These results validate \textsc{FlowQ-Net} as an efficient front-end that can feed downstream noise-aware compilation or error-mitigation pipelines with circuits that are already an order of magnitude leaner than hand-crafted or adaptive alternatives.

\subsubsection{Noisy simulation results}
To probe noise resilience we repeat the above experiments using calibrated
error profiles from publicly available IBM devices, \emph{IBM Mumbai}
\citep{patel2024curriculum} and \emph{IBM Ourense} \cite{du2022quantum}.
We compare molecular benchmarks from:  
(i) an \emph{ideal} state-vector simulation and  
(ii) noisy runs that include the calibrated error models
(iii) a single-qubit depolarizing channel with strength of $0.1 \times 10^{-2}$.
The noisy simulations are executed with built-in Pennylane noise model function with noise data obtained from previous work. 
During training, the loss is computed and optimized with respect to \emph{noisy} energy estimates obtained from these device models, and the resulting reward signal is propagated through the GFlowNet updates. Because current simulation back-ends make large-scale noisy evaluations computationally demanding, we restrict our study to the smaller molecular instances \mbox{H$_2$–2q}, \mbox{H$_2$–3q}, \mbox{H$_2$–4q}, and \mbox{LiH–4q}. For every setting, noiseless and noisy, we perform a final round of variational optimization of the continuous circuit parameters on a noiseless simulator to validate the success of maintaining chemical accuracy. %The resulting energies are reported as energy difference values to ensure fair comparison.

%For LiH-4q, whose optimization failed to reach chemical accuracy under either hardware profile. This outcome aligns with current other methods infeasible to find the circuit to the point of chemical accuracy. We also tested a synthetic single-qubit depolarizing  as follow \cite{patel2024curriculum}. 

\begin{figure}[t]
  \centering
  \includegraphics[width=.75\linewidth]{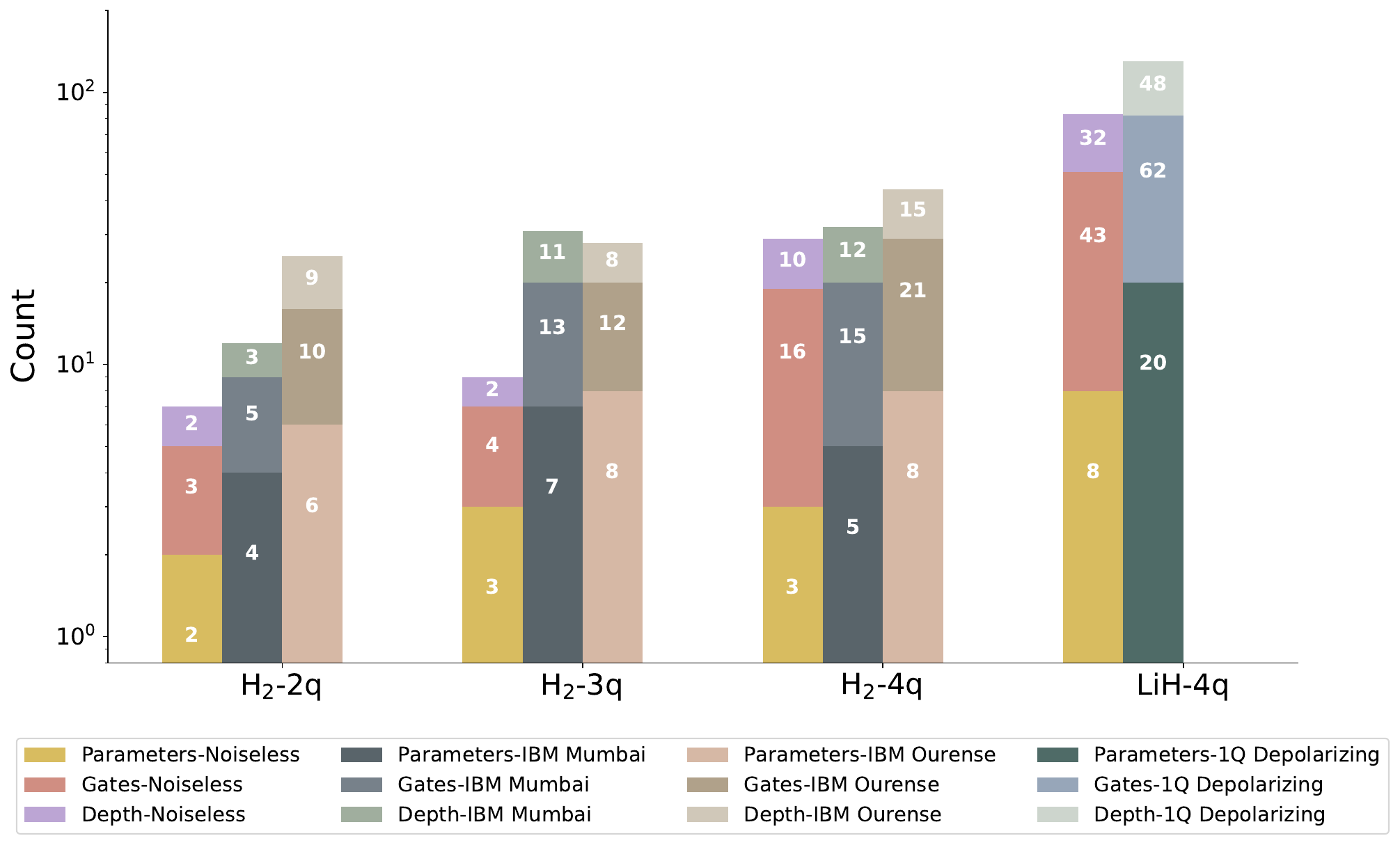}
  \caption{%
    \textbf{Quantum resource comparison across problem instances and noise models.}
    Stacked bars report the \textit{number of circuit parameters} ($P$), \textit{ gate count} ($G$), and \textit{circuit depth} ($D$) on a log-scale $y$-axis for four molecular VQE benchmarks, \(\mathrm{H}_2\) with 2, 3, and 4 qubits, and \(\mathrm{LiH}\) with 4 qubits.
    Each colour grouping corresponds to a different execution context: noiseless simulation, IBM~\textit{Mumbai} backend, IBM~\textit{Ourense} backend, and a single-qubit depolarizing noise model.
  }
  \label{fig:noise}
\end{figure}

Figure \ref{fig:noise} shows the number of trainable parameters $P$, total CNOTs $G$, and depth $D$ of the best circuit returned by \textsc{FlowQ-Net} for each molecule and noise model. For \mbox{H$_2$–2q}, \mbox{H$_2$–3q} and \mbox{H$_2$–4q}, chemical accuracy is achieved under both noise profiles at a \textit{lower gate count than UCCSD}. For LiH, chemical accuracy is achieved, but only under the depolarizing noise model.
We observe that our method returns circuits with more trainable parameters, gate counts, and greater depth as noise levels increase.  
For example, the optimal circuit for H$_2$–4q in the noiseless simulator uses only $P{=}3$, $G{=}16$, and $D{=}10$.  Under the \textit{IBM-Mumbai} profile these values rise to $P{=}5$, $G{=}15$, $D{=}12$, and they increase to $P{=}8$, $G{=}21$, $D{=}15$ with the noisier \textit{IBM-Ourense} model. The same qualitative behavior is found for the 2- and 3-qubit H$_2$ tasks, where parameter counts nearly double and depths almost triple between the ideal and noisy settings. The practical impact of this resource inflation is starkly illustrated by LiH–4q. Our methods could not find circuits achieving chemical accuracy under either hardware profile within our experimental budget. The synthetic depolarizing channel experiment for LiH-4q yielded a circuit with $P =20$, $G=62$, $D=48$ to obtain chemical accuracy. This suggests that substantially greater circuit resources are required to find optimal circuits on current noisy hardware.
%TO-DO

These findings emphasize the importance of \emph{noise-aware} circuit discovery, circuits that are optimal in silico can become sub-optimal, or even unusable, once realistic noise is considered.  By incorporating hardware models during training, \textsc{FlowQ-Net} automatically balances expressivity with robustness, but the resulting circuits may still be significantly larger than their noiseless counterparts.  Future work should therefore explore explicit multi-objective rewards that penalize depth and CNOT count more aggressively in high-noise regimes, as well as error-mitigation techniques that could help the observed resource increase.

\subsection{Image Classification with Quantum Neural Networks}
\label{sec:exp:qnn}
Next, we assess the ability of \textsc{FlowQ-Net} to discover efficient quantum neural network (QNN) architectures for image classification tasks. We follow \cite{hur2022quantum} and benchmark our method on a subset of MNIST~\cite{6296535} containing only the \texttt{0} and \texttt{1} digits. 
This yields $12\,665$ training and
$2\,115$ test images 
The image–classification task studied maps an input image $\mathbf{x}\in\mathbb{R}^{28\times28}$ to a binary label $y\in\{0,1\}$. 
We evaluate whether a \emph{parameter-efficient} quantum neural network discovered by \textsc{FlowQ-Net} can match or exceed the performance of manually designed quantum models and classical convolutional neural nets (CNNs) under identical resource constraints. 

Each image is compressed to an $m$-dimensional vector $\mathbf{x}_i=(x_i^{(1)},\dots,x_i^{(m)})$ using either principal component analysis (PCA) or a learned auto-encoder (AE); the reduced dimension is $m{=}8$ for angle encoding and $m{=}16$ for dense-angle encoding (see Appendix \ref{app:qnn} for more details). The vector is then mapped to an $m$-qubit product state $\left|\psi_0\left(x_i\right)\right\rangle$.
Given an encoding, \textsc{FlowQ-Net} explores a search space of parameterized circuits. The circuit acts on the input state to give output state $\left|\psi_t\left(\mathbf{x}_i\right)\right\rangle=\mathbf{U}_{\theta,\mathcal{G}}\left|\psi_0\left(\mathbf{x}_i\right)\right\rangle$.
And the network parameters $\boldsymbol{\theta}$ are optimized by minimizing the binary cross-entropy loss

\begin{equation}
\label{eq:ce}
    \mathcal{L}(\boldsymbol{\theta}; \mathcal{G})=\frac{1}{N} \sum_{i=1}^N\left(y_i \log \operatorname{Pr}\left(\mathcal{M}\left(\left|\psi_t\left(\mathbf{x}_i\right)\right\rangle\right)=|1\rangle\right)\right.
\left.+\left(1-y_i\right) \log \operatorname{Pr}\left(\mathcal{M}\left(\left|\psi_t\left(\mathbf{x}_i\right)\right\rangle\right)=|0\rangle\right)\right).
\end{equation}
where $\mathcal{M}$ denotes a projective measurement in the computational basis.  In the outer loop, the GFlowNet receives a reward, encouraging the discovery of shallow, high accuracy circuits. We compare against three baselines: (i) a classical CNN \cite{he2016deep}, (ii) the QCNN architecture \cite{hur2022quantum},
and (iii) QuantumDARTS \citep{wu2023quantumdarts}. 

\begin{figure}[t]
  \centering
    \includegraphics[width=\linewidth]{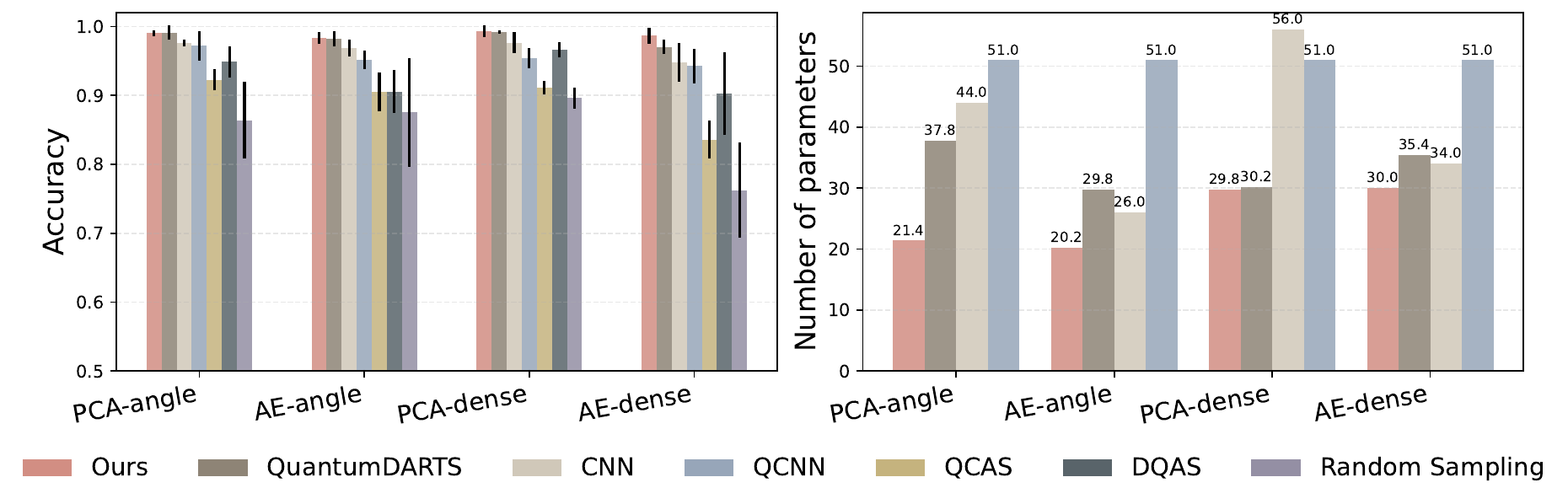}
    \caption{\textbf{Comparison of our approach with quantum and classical baselines on four data encoding setups.} Left Panel: Test accuracy. Right Panel: Average number of parameters.}
\label{fig:qnn}
\end{figure}

\paragraph{Results}
Fig. \ref{fig:qnn} lists the mean test accuracy over five runs. Across the four data–encoding scenarios our \textsc{FlowQ-Net} circuits either statistically tie or lead for the
top accuracy. They lead by $\sim\!1$–$2$ points over the best classical baseline (CNN) and by $2$–$4$ points over the widely used QCNN.  A narrow gap appears only in the \textsc{PCA–dense} case, where QuantumDARTS reaches $0.992{\pm}0.002$ versus our $0.990{\pm}0.008$, the difference falls inside standard error. \textsc{FlowQ-Net} finds circuits that are consistently shallower than the other quantum designs, as shown in Fig. \ref{fig:qnn}. Our methods require between $20.2$ and $30.0$ parameters, whereas others needs $29.8$–$56.0$ parameters, which provide a $25$–$45\%$ reduction.

The experiment confirms that generative, reward-guided circuit search can discover compact image classification QNNs that \emph{simultaneously} achieve leading accuracy and reductions in parameter count, two key points for deployment on near-term quantum hardware.

\subsection{Combinatorial-optimization: Max-Cut}
\label{sec:exp:maxcut}
Next, we benchmark on unweighted Max-Cut problem for graphs generated by the Erdős–Rényi model with 10 vertices at varying edge creation probabilities ($p_e$). See Appendix \ref{app:maxcut} for problem setting and experimental details. 

\begin{table}[ht]
  \small
  \setlength{\tabcolsep}{2.5pt}
  \caption{Unweighted Max-Cut on Erdős–Rényi graphs with 10 vertices (ten random graphs per edge-creation probability $p_e$). Each block reports the edge count $|E|$, mean degree $\overline{D}$, minimum degree $D_{\min}$, maximum degree $D_{\max}$, and the mean circuit depth.  QuantumDARTS does not publish mean depths; the depths shown here come from their exemplar solutions.}
  \label{tab:mc}
  \centering
    \resizebox{\linewidth}{!}{%
    \begin{tabular}{lcccccccccc}
    \toprule
      & \multicolumn{5}{c}{\textbf{ \textsc{FlowQ-Net} (Ours})} & \multicolumn{5}{c}{\textbf{QuantumDARTS}} \\
      \cmidrule(lr){2-6}\cmidrule(lr){7-11}
      $p_e$ &
      $|E|$ & $\overline{D}$ & $D_{\text{min}}$ & $D_{\text{max}}$ & \textbf{Depth}$\!\downarrow$ &
      $|E|$ & $\overline{D}$ & $D_{\text{min}}$ & $D_{\text{max}}$ & \textbf{Depth}$\!\downarrow$ \\
    \midrule
    0.25 &
      12.80 $\pm$ 2.70 & 2.56 $\pm$ 0.54 & 0.70 $\pm$ 0.67 & 4.40 $\pm$ 0.70 & 20.2 $\pm$ 4.9 &
      12.6  $\pm$ 3.0  & 2.5  $\pm$ 1.3  & 0.6  $\pm$ 0.5  & 4.6  $\pm$ 1.4 & 30 \\
    0.50 &
      22.20 $\pm$ 3.05 & 4.44 $\pm$ 0.61 & 2.10 $\pm$ 0.99 & 6.50 $\pm$ 0.71 & 24.1 $\pm$ 10.6 &
      22.1  $\pm$ 3.8  & 4.4  $\pm$ 1.6  & 2.5  $\pm$ 1.0  & 6.6  $\pm$ 0.8 & 33 \\
    0.75 &
      34.20 $\pm$ 4.10 & 6.84 $\pm$ 0.82 & 4.60 $\pm$ 1.35 & 8.50 $\pm$ 0.70 & 26.8 $\pm$ 10.3 &
      33.5  $\pm$ 3.1  & 6.7  $\pm$ 1.4  & 5.0  $\pm$ 0.9  & 8.4  $\pm$ 0.5 & 34 \\
    \bottomrule
  \end{tabular}}
\end{table}

The mean values of key metrics such as the number of edges ($N_\text{edges}$), mean degree (D), minimum degree ($D_{min}$), and maximum degree ($D_{max}$) were reported, along with the Conditional Value-at-Risk (CVaR) metric and the depth of the solution quantum circuits. 

We report the statistics of our results in Table \ref{tab:mc}. We compared our results with those from other works, noting that only QuantumDARTS \cite{wu2023quantumdarts} reports successfully solving the 10-node unweighted Max-Cut problem. However, our method produces more efficient circuits than those reported by QuantumDARTS. On average, our model discovers quantum circuits with depths of 20.20, 24.10, and 26.80 for the respective $p_e$ values, compared to depths of 30, 33, and 34 in their results. This demonstrates that our approach is capable of generating more compact circuits, highlighting its efficiency. 

\begin{figure}[htbp]
  \centering
  \begin{subfigure}{.3\textwidth}
    \centering
    \includegraphics[scale=.2]{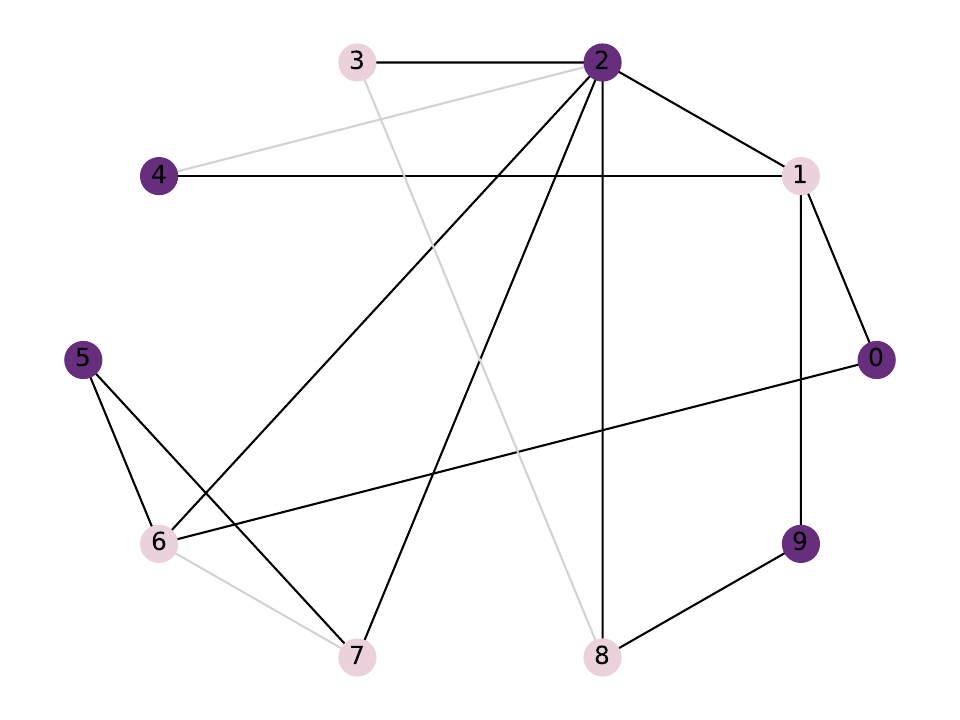}
    \caption{$p_e = 0.25$}
    \label{fig:mc_25}
  \end{subfigure}%
  \begin{subfigure}{.3\textwidth}
    \centering
    \includegraphics[scale=.2]{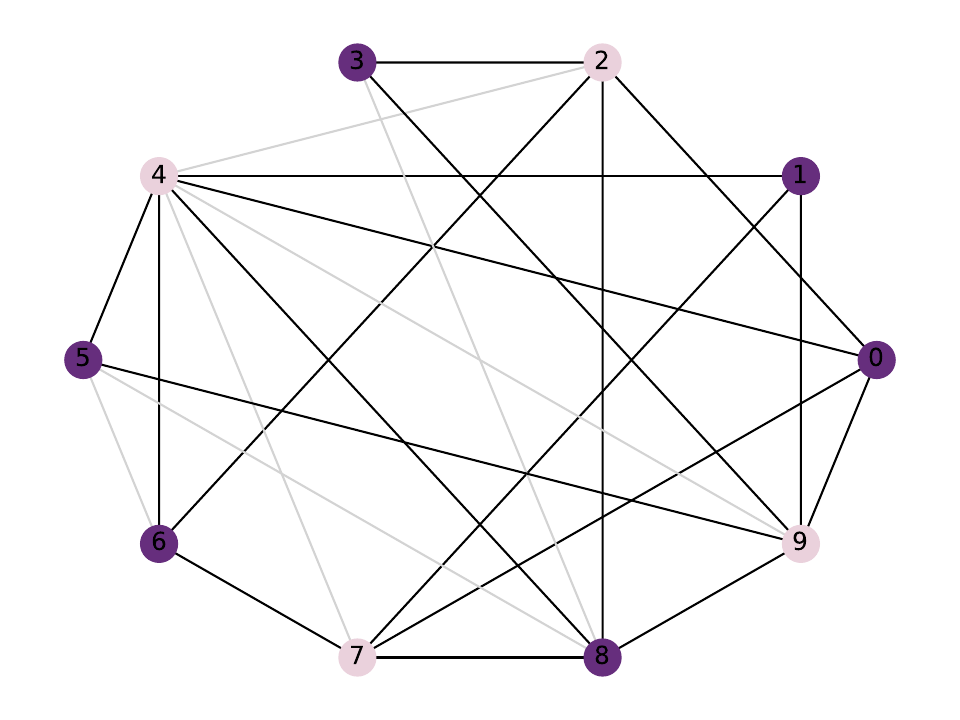}
    \caption{$p_e = 0.5$}
    \label{fig:mc_50}
  \end{subfigure}
  \begin{subfigure}{.3\textwidth}
    \centering
    \includegraphics[scale=.2]{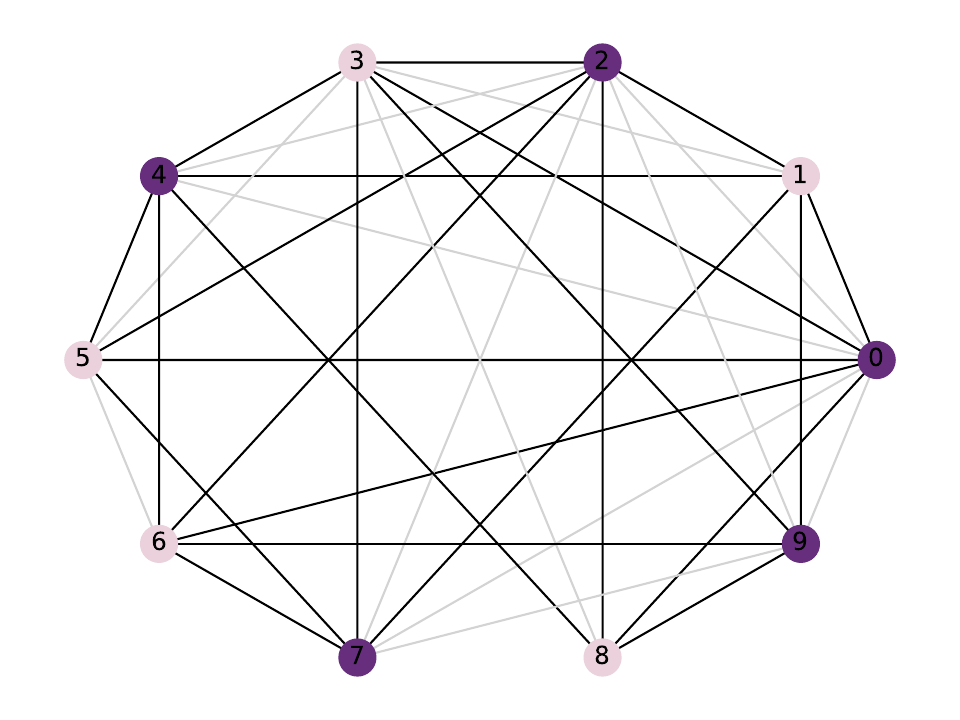}
    \caption{$p_e = 0.75$}
    \label{fig:mc_75}
  \end{subfigure}
  \caption{Optimal Max-Cut solutions. Vertices are split into light-purple and deep-purple sets. Black edges cross the cut, light-gray edges lie within a set.}
 \label{fig:mc}
\end{figure}

Figure \ref{fig:mc} shows the visual representations of these solutions for the probabilities of edge creation $p_e = 0.25$, $p_e = 0.5$, and $p_e = 0.75$, respectively, depict the division of vertices into two sets, with edges between sets shown in black and within sets in light gray. This clear distinction in the graph's structure highlights the effectiveness of GFlowNets in finding optimal cut solutions at different graph densities. A details of the Max-Cut benchmarks, including QAOA and Goemans–Williamson comparisons, is provided in Appendix~\ref{app:maxcut}, confirming the consistent advantage and circuit efficiency of \textsc{FlowQ-Net} across varying graph densities.

\section{Conclusion}
\label{sec:con}

We introduced \textsc{FlowQ-Net}, a generative framework that reformulates quantum-circuit synthesis as a sequential decision process solved by Generative Flow Networks. By sampling circuits in proportion to a flexible reward, \textsc{FlowQ-Net} learns to generate a diverse family of high-quality designs rather than converging on a single optimum. This diversity enables the discovery of compact, high-performing architectures that use up to an order of magnitude fewer gates than conventional approaches—a promising step toward resource-efficient quantum algorithms.

Across four molecular VQE benchmarks, circuits discovered by \textsc{FlowQ-Net} preserved chemical accuracy while reducing parameter counts, depths, and gate numbers by up to an order of magnitude compared with UCCSD, ADAPT-VQE, QuantumDARTS, and CRLQAS. When realistic noise models from the IBM \textit{Mumbai} and \textit{Ourense} back-ends were incorporated during training, the framework adapted by producing deeper yet still competitive circuits, demonstrating its capability for noise-aware optimization. Beyond quantum chemistry, experiments on unweighted Max-Cut and binary image classification showed that \textsc{FlowQ-Net} can design shallow, high-performing circuits for combinatorial optimization and quantum neural networks, achieving accuracy comparable to or exceeding classical and quantum baselines while using substantially fewer resources.

Overall, these results establish that diversity-seeking generative policies provide a principled route to resource-efficient and noise-robust quantum algorithms. Future work will involve deploying the generated circuits on real hardware to quantify fidelity gains and extending the reward formulation to incorporate explicit multi-objective and error-mitigation criteria. Applying the framework to tasks such as unitary compilation, entangled-state generation, and quantum error-correcting code synthesis will further test its versatility and may accelerate the automated discovery of practical circuits for the NISQ era and beyond.

\bibliographystyle{ACM-Reference-Format}
\bibliography{reference}
\newpage
\appendix

\section{Implementation Details of \textsc{FlowQ-Net}}
\label{app:imple}

The routine in Algorithm~\ref{alg:gfn_qas} alternates between \emph{generation} and \emph{policy update}.
During each update round we roll out a mini‑batch of $B$ trajectories with the current forward flow model $P_{F,\phi}$.
Beginning from the empty sequence $s_0$, the policy selects gates one at a time,subject to the forward mask in
Appendix~\ref{apx:mask},until the trajectory reaches a terminal state or the maximum depth $d_{\max}$.
For every completed circuit $s$, an inner VQA loop optimizes the continuous parameters $\boldsymbol{\theta}$ and returns the task‑loss $\mathcal{L}_{\mathcal{T}}(s,\boldsymbol{\theta})$.
We convert this loss to a non‑negative reward $R = \exp\!\bigl[-\beta\bigl(\mathcal{L}_{\mathcal{T}}
(s,\boldsymbol{\theta}) - b\bigr)\bigr],$ store $(\tau,R)$ in the mini‑batch, and track the best $(\mathcal{G}^{\star},\boldsymbol{\theta}^{\star})$ encountered so far.

Once $B$ trajectories are collected, we perform a single gradient step on $\phi$ and the normalising scalar $Z_\phi$ by minimising the trajectory‑balance objective averaged over the batch.
Because all samples are on‑policy, no replay buffer or importance weights are required, which simplifies implementation and eliminates off‑policy instabilities. The forward mask (see appendix \ref{apx:mask}) guarantees syntactic validity and removes redundant actions. Together these constraints enable fast, stable training and yield compact circuits that satisfy hardware depth limits, as confirmed by the empirical results in Section~\ref{sec:exp}.

\begin{algorithm}[h]
\caption{\textsc{FlowQ-Net}}
\label{alg:gfn_qas}
\KwIn{Task $\mathcal{T}$, gate set $\mathbf{G}$, inverse temperature $\beta$,  
      max depth $d_{\max}$, max \# rotation gates $p_{\max}$,  
      batch size $B$, learning rates $\eta_{F},\eta_{Z}$,  
      total updates $U$.}
\KwOut{Best circuit $\mathcal{G}^{\star}$ and parameters $\boldsymbol{\theta}^{\star}$.}

\smallskip
\textbf{Initialization:}\;  
Forward policy $P_{F,\phi}$ (Transformer), partition scalar $Z_\phi$,  
best loss $\mathcal{L}^{\star}\!\leftarrow\!+\infty$.

\BlankLine
\For{update $u \leftarrow 1$ \KwTo $U$}{%
  \tcp{Collect a batch of $B$ trajectories on‑policy}
  \For{$i \leftarrow 1$ \KwTo $B$}{%
     $s \leftarrow s_{0}$;\; $\tau \leftarrow \emptyset$;

     \While{$s$ non‑terminal \textbf{and} depth$(s)<d_{\max}$}{%
        Compute mask $\mathcal{M}(s)$ (App.~\ref{apx:mask});\\
        Sample $a \sim P_{F,\phi}(\cdot\mid s)$ with mask $\mathcal{M}(s)$;\\
        $s' \leftarrow \textsc{ApplyGate}(s,a)$;\;
        $\tau \leftarrow \tau \cup \{(s,a,s')\}$;\;
        $s \leftarrow s'$;%
     }

     \tcc{Inner VQA loop to score the sampled circuit}
     $\boldsymbol{\theta}^{\star}(s)
        \leftarrow \arg\min_{\boldsymbol{\theta}}
        \mathcal{L}_{\mathcal{T}}(s,\boldsymbol{\theta})$;\\
     $R_i \leftarrow
        \exp\!\bigl[-\beta\bigl(\mathcal{L}_{\mathcal{T}}
        (s,\boldsymbol{\theta}^{\star}(s)) - b\bigr)\bigr]$;\\
     Store $(\tau_i,R_i)$ in the mini‑batch $\mathcal{B}$;\\[2pt]
     \If{$\mathcal{L}_{\mathcal{T}}(s,\boldsymbol{\theta}^{\star}(s)) < \mathcal{L}^{\star}$}{%
         $\mathcal{L}^{\star}\!\leftarrow\!
           \mathcal{L}_{\mathcal{T}}(s,\boldsymbol{\theta}^{\star}(s))$;\\
         $\mathcal{G}^{\star}\!\leftarrow s$;\;
         $\boldsymbol{\theta}^{\star}\!\leftarrow
           \boldsymbol{\theta}^{\star}(s)$;%
     }%
  } % end for i

  \tcc{Trajectory‑balance gradient step (Eq.~\ref{eq:tb})}
  Compute $\mathcal{L}_{TB}(\mathcal{B};\phi)$ as the mean over batch;\\
  $\phi \leftarrow \phi - \eta_{F}\nabla_{\phi}\mathcal{L}_{TB}$;\quad
  $Z_{\phi} \leftarrow Z_{\phi} - \eta_{Z}\partial_{Z}\mathcal{L}_{TB}$;\\
  Clear batch $\mathcal{B}$;
} % end for u

\Return{$(\mathcal{G}^{\star},\boldsymbol{\theta}^{\star})$.}
\end{algorithm}

\paragraph{Forward Mask}
\label{apx:mask}
To enforce constraints on the gate selection process, we define a mask that evaluates the feasibility of applying each quantum gate based on the sequence of previous operations. The mask prevents the model from selecting invalid or redundant actions. Specifically, we define the following rules: (1) A gate cannot be applied consecutively to the same qubit, meaning the selected gate must differ from the previous gate acting on that qubit. (2) Rotation gates from the set ${\mathrm{R_X},\mathrm{R_Y}, \mathrm{R_Z}}$ are not allowed to follow another rotation gate, ensuring a diverse set of operations. (3) A two-qubit gate, $\mathrm{CNOT}$, can only be applied if at least one other gate has already been applied to one of the involved qubits, ensuring that multi-qubit operations are meaningful within the context of the quantum state evolution. (4) The model should avoid redundant two-qubit gates, such as applying a $\mathrm{CNOT}$ with control on qubit 0 and target on qubit 1 right after a $\mathrm{CNOT}$ with control on qubit 1 and target on qubit 0.

\section{Supplements to quantum chemistry}
\subsection{ Dynamic Lie algebra generator for H$_2$-4q}
\label{sec:dla}

\begin{figure}[hb]
  \centering
  \includegraphics[width=0.85\linewidth]{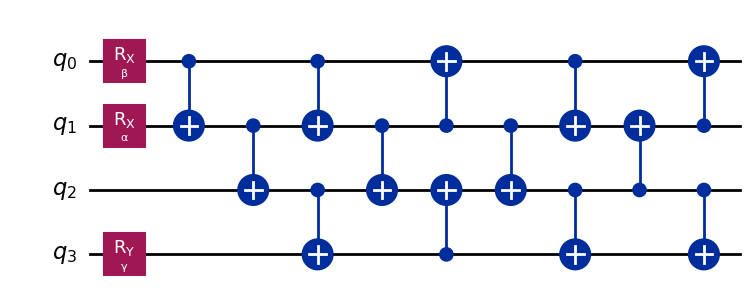}
  \caption{\textbf{Example circuit discovered by \textsc{FlowQ-Net} for the H$_2$–4q system.} 
  The ansatz consists of entangling CNOT layers (blue) interleaved with single-qubit rotation gates (red). 
  The resulting architecture contains three variational parameters $(\alpha, \beta, \gamma)$ and achieves chemical accuracy while remaining substantially shallower than standard UCCSD constructions.}
  \Description{Circuit diagram of the FlowQ-Net-generated H2-4q ansatz showing CNOT entangling layers and three single-qubit rotation gates labeled R_X and R_Y.}
  \label{fig:h2_circuit}
\end{figure}
To further assess the expressivity of the discovered ansatz, we conducted a Dynamical Lie Algebra (DLA) \cite{schirmer2002identification,diaz2023showcasing} analysis on the H$_2$–4q system. The objective is to verify whether the trained circuit generates the minimal algebra required to span the molecular ground‐state manifold.

We first consider the logical two‐level subspace $\mathcal{H}_{\text{pair}} = \mathrm{span}\{\ket{0_L},\ket{1_L}\}$, where $\ket{0_L}=\ket{0011}$ and $\ket{1_L}=\ket{1100}$ correspond to the dominant configurations of the molecular Hamiltonian. The projector onto this subspace is denoted by $P$. Defining the logical Pauli operator
\[
J_y^{(L)} \equiv i\big(\ket{1_L}\!\bra{0_L} - \ket{0_L}\!\bra{1_L}\big),
\]
we obtain the single‐parameter family
\[
e^{-i\theta J_y^{(L)}}\ket{0_L}
  = \cos\theta\,\ket{0_L}+\sin\theta\,\ket{1_L},
\]
which represents the exact ground‐state manifold of H$_2$. Consequently, the corresponding DLA is
\(
\mathfrak{L}_{\text{state}} = \mathrm{span}\{\, iJ_y^{(L)} \,\}.
\)

To identify its physical realization within the four‐qubit Hilbert space, we evaluate the action of the Pauli string $Y_3X_2X_1X_0$ on the logical basis,
\[
Y_3X_2X_1X_0\ket{0011}= i\ket{1100},\qquad
Y_3X_2X_1X_0\ket{1100}= -i\ket{0011}.
\]
Hence, projection onto $\mathcal{H}_{\text{pair}}$ yields
\[
P(Y_3X_2X_1X_0)P = \sigma_y \equiv J_y^{(L)},
\]
showing that the single physical generator $iY_3X_2X_1X_0$ exactly reproduces the logical transformation required to generate the ground‐state manifold.

The trained circuit $U(\boldsymbol{\theta})$ contains three continuous parameters associated with rotations $R_x^{(1)}(\alpha)$, $R_x^{(0)}(\beta)$, and $R_y^{(3)}(\gamma)$, followed by a sequence of CNOT gates.
Conjugating the rotation generators by $T$ gives the effective output‐frame generators
\[
\widetilde{X}_1 = X_3 X_0,\qquad
\widetilde{X}_0 = X_3 X_1,\qquad
\widetilde{Y}_3 = Y_3X_2X_1X_0.
\]
These generators mutually commute,
\[
[X_3X_0,X_3X_1]=[X_3X_0,Y_3X_2X_1X_0]=[X_3X_1,Y_3X_2X_1X_0]=0,
\]
so that the circuit’s full‐space algebra is
\[
\mathfrak{L}(U)=\mathrm{span}\{\,iX_3X_1,\ iX_3X_0,\ i(Y_3X_2X_1X_0)\,\},
\]
an abelian subalgebra of dimension three in $\mathfrak{su}(16)$. Projecting onto $\mathcal{H}_{\text{pair}}$ gives
\[
P\widetilde{X}_1P = P\widetilde{X}_0P = 0,\qquad P(Y_3X_2X_1X_0)P = \sigma_y,
\]
and therefore the logical DLA of the circuit reduces to
\[
\mathfrak{L}_L(U)=P\,\mathfrak{L}(U)\,P=\mathrm{span}\{\, i\sigma_y \,\}.
\]
This is identical to the one‐dimensional algebra obtained from the state‐manifold analysis, indicating that the learned circuit generates precisely the transformation
\[
e^{-i\gamma\sigma_y}\ket{0_L}
  = \cos\gamma\,\ket{0_L}+\sin\gamma\,\ket{1_L},
\]
where $\gamma=2\theta$ under the convention $R_y(\gamma)=e^{-i\gamma Y/2}$. 

The analysis confirms that the circuit discovered by \textsc{FlowQ-Net} faithfully realizes the minimal Lie algebra $\mathfrak{L}_L(U)=\mathrm{span}\{i\sigma_y\}$ corresponding to the physical generator $iY_3X_2X_1X_0$. This generator forms the entangling transformation connecting the Hartree–Fock reference to the exact ground state, thereby validating that the algorithm learns the physically relevant subalgebra required for accurate state preparation while maintaining a compact, resource‐efficient structure.

\section{Supplements to Image Classification Benchmarks}
\label{app:qnn}

In our image classification benchmarks, we follow the experimental protocols established in Ref~\cite{hur2022quantum}. The goal is to evaluate quantum neural network architectures on standard image datasets with reduced data dimensionality and efficient quantum encoding. 

To manage high-dimensional image data, we employ a classical autoencoder as a preprocessing step. The autoencoder reduces each image to a compact latent representation, which is then fed into the quantum circuit. Following Ref~\cite{hur2022quantum}, both encoder and decoder are implemented as fully connected neural network layers with ReLU and sigmoid activations, respectively. The hidden dimension is set to 8 for simple angle encoding and 16 for dense angle encoding. The autoencoder is trained on the training set for 10 epochs using the mean squared error loss and optimized using Adam.

To efficiently encode the reduced data into quantum circuits, we use two schemes: angle encoding and dense angle encoding, chosen for their low gate count. 

\paragraph{Angle encoding.}
Given a latent vector  $x_i = (x_i^{(1)}, \ldots, x_i^{(m)})$, the quantum state is prepared as

$$
\left|\psi\left(x_i\right)\right\rangle=\bigotimes_{j=1}^m\left(R_y\left(x_i^{(j)}\right)|0\rangle\right) .
$$

\paragraph{Dense angle encoding.}
For $m$, we use a more compact encoding,
$$
\left|\psi\left(x_i\right)\right\rangle=\bigotimes_{j=1}^{m / 2}\left(R_y\left(x_i^{(j+m / 2)}\right) R_x\left(x_i^{(j)}\right)|0\rangle\right) .
$$

For classification, the circuit output is measured, and the probability of a particular class is computed by summing the probabilities of all basis states corresponding to that class. For binary classification, the probability $P(\mathcal{M}(|\psi_t(x_i)\rangle) = 0)$ is the sum over all even-parity basis states (e.g., $|000\rangle$, $|010\rangle$, etc.), while $P(\mathcal{M}(|\psi_t(x_i)\rangle) = 1)$ is the sum over all odd-parity basis states (e.g., $|001\rangle$, $|011\rangle$, etc.). The two probabilities are complementary and sum to one.

\section{Supplements to Combinatorial-optimization: Max-Cut}
\label{app:maxcut}

\begin{table}[t]
\centering
\small
\setlength{\tabcolsep}{5pt}
\renewcommand{\arraystretch}{1.1}
\caption{%
Comparison of \textsc{FlowQ-Net}, QAOA ($p{=}2$) with two optimizers, 
and the classical Goemans–Williamson (GW) algorithm 
on 10-node Erdős–Rényi Max-Cut graphs. 
Values are mean cut ratios over 10 random graphs per edge-creation probability $p_e$.
}
\label{tab:maxcut_qaoa_gw}
\vspace{0.4em}
\begin{tabular}{lcccc}
\toprule
\textbf{$p_e$} &
\textbf{FlowQ-Net (Ours)} &
\textbf{QAOA–Adam} &
\textbf{QAOA–COBYLA} &
\textbf{GW (classical)} \\
\midrule
0.25 & \textbf{0.994 $\pm$ 0.012} & 0.724 $\pm$ 0.102 & 0.779 $\pm$ 0.089 & 0.878 \\
0.50 & \textbf{0.990 $\pm$ 0.020} & 0.760 $\pm$ 0.056 & 0.784 $\pm$ 0.057 & 0.878 \\
0.75 & \textbf{0.985 $\pm$ 0.041} & 0.833 $\pm$ 0.025 & 0.805 $\pm$ 0.036 & 0.878 \\
\bottomrule
\end{tabular}
\Description{%
Table comparing mean Max-Cut ratios for FlowQ-Net, QAOA with Adam and COBYLA optimizers, and the classical Goemans–Williamson method.
FlowQ-Net consistently achieves higher mean cut ratios across all edge densities while using shallower circuits.%
}
\end{table}

We benchmark on unweighted Max-Cut for graphs generated by the Erdős–Rényi model with 10 vertices at varying edge creation probabilities ($p_e$). Problem Hamiltonians are encoded via the standard Ising formulation \cite{farhi2014quantum} and mapped to qubits.  The observables can be calculated as
\begin{equation}
\mathcal{O}_c=\sum_{e_{i, j} \in \mathcal{E}} \frac{1}{2}\left(I-Z_{i} Z_{j}\right),
\end{equation}
where $I$ is the identity matrix and $Z_i$ is the Pauli-Z matrix acting on qubit $i$. The eigenstate with the largest eigenvalue represents the solution to the Max-Cut problem. 

For all benchmarks, problem instances are generated using fixed random seeds for reproducibility. Each Hamiltonian is constructed directly from the adjacency matrix of the graph, ensuring a one-to-one correspondence between the graph structure and the quantum observable. 

The results in Table~\ref{tab:maxcut_qaoa_gw} highlight the strong performance of \textsc{FlowQ-Net} on combinatorial optimization tasks. Across all edge-creation probabilities, our method consistently achieves the highest mean cut ratios, exceeding both quantum and classical baselines. While the Goemans–Williamson algorithm\cite{goemans1995improved} provides a classical upper reference with a constant approximation ratio of $0.878$, \textsc{FlowQ-Net} surpasses this bound on average, demonstrating that generative circuit design can identify problem-adapted ansätze that effectively exploit quantum interference to capture optimal cuts. In contrast, QAOA \cite{choi2019tutorial} with either Adam or COBYLA optimization yields lower cut ratios and exhibits higher variance, reflecting its sensitivity to initialization and local minima. Notably, the improvement of \textsc{FlowQ-Net} becomes more pronounced as the graph density increases ($p_e{=}0.75$), where the search space is more complex. These findings validate that reward-guided generative exploration enables our framework to discover compact, high-performing quantum circuits that generalize across varying graph topologies.

\end{document}